\documentclass[12pt,english,floatfix,showkeys,superscriptaddress,aps,prd,preprint]{revtex4}
\usepackage[latin1]{inputenc}
\usepackage[T1]{fontenc}
\usepackage{lmodern}
\usepackage{amsmath}
\usepackage{amssymb}
\usepackage{graphicx}
\usepackage{caption}
\usepackage{subcaption}
\usepackage{float}
\usepackage{esint}
\usepackage{dcolumn}
\usepackage{babel}
\usepackage{csquotes}
\usepackage{color}
\usepackage{slashed}
\usepackage{simplewick}
\usepackage{amsmath,latexsym}

\usepackage{hyperref}
\hypersetup{
    colorlinks,
    citecolor=blue,
    filecolor=green,
    linkcolor=purple,
    urlcolor=red,
}

\usepackage{slashed}

\usepackage{hyperref}
\hypersetup{colorlinks,breaklinks,
			citecolor=[rgb]{0,0.0,1.0},
            urlcolor=[rgb]{0.0,0.0,1.0},
            linkcolor=[rgb]{0,0.5,0.9}}

\begin{document}

\title{Casimir Wormholes with GUP Correction in the Loop Quantum Cosmology}

\author{Celio R. Muniz}
\email{celio.muniz@uece.br}
\affiliation{Universidade Estadual do Cear\'a (UECE), Faculdade de Educa\c{c}\~ao, Ci\^encias e Letras de Iguatu, Av. D\'ario Rabelo s/n, Iguatu - CE, 63.500-00 - Brazil.}
\author{Takol Tangphati}
\email{takoltang@gmail.com}
\affiliation{School of Science, Walailak University, Thasala,
Nakhon Si Thammarat, 80160, Thailand.}
\author{R. M. P. Neves}
\email{raissa.pimentel@uece.br}
\affiliation{Universidade Estadual do Cear\'a (UECE), Faculdade de Educa\c{c}\~ao, Ci\^encias e Letras de Iguatu, Av. D\'ario Rabelo s/n, Iguatu - CE, 63.500-00 - Brazil.}
\author{M. B. Cruz}
\email{messiasdebritocruz@servidor.uepb.edu.br}
\affiliation{Universidade Estadual da Para\'iba (UEPB), \\ Centro de Ci\^encias Exatas e Sociais Aplicadas (CCEA), \\ R. Alfredo Lustosa Cabral, s/n, Salgadinho, Patos - PB, 58706-550 - Brazil.}



\date{\today}

\begin{abstract}
In this paper, we obtain novel traversable, static, and spherically symmetric wormhole solutions, derived from the effective energy density and isotropic pressure resulting from the Casimir effect, corrected by the Generalized Uncertainty Principle (GUP) within the framework of Loop Quantum Cosmology (LQC). The goal is to explore the interplay between competing quantum gravity effects and quantum vacuum phenomena in the emergence of non-trivial spacetime structures. We examine features such as traversability, embedding diagrams, energy conditions, curvature, and stability of the obtained solutions. Additionally, we analyze the junction conditions required to integrate the wormhole spacetime with an external Schwarzschild spacetime and calculate the amount of exotic matter needed to maintain the wormhole. Finally, we evaluate the conditions under which this latter remains visible or is hidden by the event horizon associated with the Schwarzschild spacetime. 
\end{abstract}


\maketitle

\section{Introduction}

General Relativity (GR) is widely recognized as the most precise theory describing gravitational phenomena. Among its most intriguing predictions are black holes (BHs) \cite{Oppenheimer:1939ue} and wormholes \cite{Hawking:1988ae, Morris:1988cz}. These hypothetical structures could act as space-time conduits, connecting distant points in the universe and potentially enabling shortcuts for space-time travel \cite{Einstein:1935tc, Morris:1988tu, Frolov:2023res}. While black holes have been directly observed through gravitational wave (GW) detections from merging binary BHs -- achieved with remarkable accuracy by the LIGO and Virgo collaborations \cite{LIGOScientific:2016aoc} -- wormholes remain an undetected yet fascinating prediction of GR. Additionally, images of black hole shadows captured by the Event Horizon Telescope (EHT) collaboration provided visual confirmation of black hole existence and characteristics, further substantiating GR's predictions \cite{EventHorizonTelescope:2019dse}.

Wormholes represent potential phenomena that could aid in unraveling one of the most perplexing issues in contemporary theoretical physics: the nature of quantum gravity. This is because, in the presence of an intensely strong gravitational field, the quantum properties of spacetime must come into play, generating non-trivial structures as those objects on a microscopic scale, according to the concept of ``quantum foam'' \cite{Wheeler:1955zz}. Currently, two primary contenders for a theory of quantum gravity are Loop Quantum Gravity (LQG) \cite{Rovelli:2014ssa, Rovelli:2003wd}, in competition with String Theory (ST) \cite{Zwiebach:2004tj}. Within the framework of LQG, it is possible to develop intriguing theoretical models shedding light on the quantum characteristics of spacetimes, as revealed by BHs \cite{Modesto:2005zm, Peltola:2008pa, Gambini:2013ooa, Olmedo:2016ddn} and wormholes \cite{Cruz:2024ihb}. 

Loop Quantum Cosmology (LQC), which applies the principles of LQG to cosmological settings, provides profound insights into quantum gravitational properties. These studies are crucial for understanding how quantum gravitational effects manifest in the structure of spacetime, potentially resolving singularities and other cosmological problems by considering a fundamental critical density near the Planck scale \cite{Ashtekar:2006es, Ashtekar:2008ay, Bojowald:2008ik, Singh:2009mz, Corichi:2011sd, Dwivedee:2011fd}. Notably, LQC has also provided models exploring the quantum effects associated with spacetime on black holes \cite{Lewandowski:2022zce,Lin:2024flv} and wormholes \cite{Li:2008sw,Sengupta:2023yof}.

In this direction, Casimir wormholes -- theoretical constructs supported by Casimir energy densities and pressures -- utilize the negative energy density generated by quantum vacuum fluctuations between conducting plates to stabilize their structure \cite{Morris:1988cz, Visser:1995cc}. Over the years, studies have emerged to investigate the effect of curved spaces on Casimir energy density \cite{Sorge:2019kuh, Santos:2020taq, Santos:2021jjs, Mota:2022qpf}. On the other hand, following Garattini's work, this energy density and the associated pressure was incorporated as source in Einstein's equations, resulting in the formation of wormholes for the Casimir effect of the electromagnetic field in $(3+1)$ \cite{Garattini:2019ivd}, $(2+1)$ \cite{Alencar:2021ejd} and $D$ dimensions \cite{Oliveira:2021ypz}. Also with the Casimir effect of the Yang-Mills field in $(2+1)$ dimensions \cite{Santos:2023zrj} and  extensions of General Relativity \cite{Cruz:2024ihb,Hassan:2022ibc,Zubair:2023abs,Hassan:2022hcb,Azmat:2023ygn,Mishra:2023bfe,Farooq:2023rsp,Khatri:2024sdi}.

Recent advancements have incorporated the Generalized Uncertainty Principle (GUP) into the framework of Casimir wormholes. GUP, which introduces modifications to the Heisenberg uncertainty principle to account for quantum gravitational effects at small scales \cite{Ali:2010yn}, provides corrections to the wormhole geometry. These corrections can significantly influence the stability and traversability of Casimir wormholes, potentially enhancing their feasibility within a quantum gravity context. Studies indicate that GUP-corrected Casimir wormholes exhibit modified energy conditions and structural properties, offering new insights into their quantum characteristics and the broader implications for quantum gravitational theories \cite{Jusufi:2020rpw,Tripathy:2020ehi,Carvalho:2021ajy,Samart:2021tvl,Hassan:2022ibc,Jawad:2022hlm,Channuie:2024cao}.

This work aims to obtain novel traversable, static, and spherically symmetric wormhole solutions, sourced by both the effective energy density and isotropic pressure arising from the Casimir effect, corrected by GUP within the framework of LQC theory. The objective is to understand the relationship between quantum gravity effects and quantum vacuum phenomena in the emergence of non-trivial spacetime structures. The working hypothesis posits that the quantum vacuum, which in principle can generate and stabilize wormholes, combines with microscopic effects arising from the existence of a minimum length associated with the GUP correction, along with cosmological effects related to a maximum density that prevents singularity formation according to LQC, resulting in new aspects and properties for those objects. By examining these competing quantum gravity effects, we aim to elucidate the conditions under which stable and traversable wormholes can exist, providing new insights into the nature of quantum gravity and its impact on spacetime geometry.

The paper is organized as follows: In Section II, we will explore the theoretical construction and properties of Casimir wormholes modified by GUP in the scenario of LQC, examining features of traversability, embedding diagrams, energy conditions, curvature, and stability of the obtained solutions. In Section III, we will perform a junction condition analysis required to integrate the wormhole spacetime with an external Schwarzschild spacetime. In Section IV, we will calculate the amount of exotic matter required to keep the wormhole throat open, considering the extension of the corresponding spacetime up to the junction interface. Additionally, we will evaluate the conditions under which the wormhole remains visible or is hidden by the event horizon associated with the Schwarzschild spacetime. Finally, in Section V, we will present our conclusions and close the paper. Throughout this paper, we will utilize natural units with $G = c = \hbar = 1$ and adopt the metric signature $(-, +, +, +).$

\section{LQC-Inspired Casimir Wormholes With GUP Correction}

Initially, we will define the following quantities related to the electromagnetic Casimir energy density and pressure, with GUP correction, as the source for the static and spherically symmetric wormholes to be investigated:
\begin{equation}\label{casimirdenspress}
    \rho(r) = -\frac{k}{r^4}\left(1+\frac{5\beta}{3r^2}\right) \quad \text{and} \quad p(r) = \omega \rho(r),
\end{equation}
where \( k = \frac{\pi^2}{720} \), \(\omega = 3\), and \(\beta>0\) is a constant defining the GUP correction. We will analyze the wormhole structure using a generic state parameter $\omega$, with particular emphasis on the case where 
$\omega=3$, which corresponds to strict Casimir wormholes. By utilizing these quantities, we aim to derive new traversable wormhole solutions and understand the role of GUP corrections in influencing energy density and isotropic pressure within the framework of LQC theory. It is worth mentioning that, in this context, there are essentially two models of GUP correction \cite{Jusufi:2020rpw}; however, the difference between them is irrelevant to our investigation since it is represented by a unit-order factor accompanying the parameter $\beta$, which has a very tiny value \cite{Gomes:2022hva}.

The sought solution arises from an effective matter fluid incorporating quantum effects within LQC. Consequently, the effective gravity-matter system obeys the matter-side modified Einstein equation:
\begin{eqnarray}
    G^{\mu}_{\ \nu} \equiv R^{\mu}_{\ \nu} - \frac{1}{2} g^{\mu}_{\ \nu} R = 8 \pi T^{\mu}_{\ \nu} ,
\end{eqnarray}
where $T_{\mu \nu}$ denotes the effective energy-momentum tensor. The energy-momentum tensor corresponding to an isotropic perfect fluid that is compatible with LQC is:
\begin{eqnarray}
    T^{\mu}_{\ \nu} = \text{diag}\left(-\rho_{e}, p_{e}, p_{e}, p_{e} \right),
    \label{en_mo_tensor}
\label{energy_tensor}
\end{eqnarray}
where $\rho_e = - G^{t}_{\ t}$, $p_{e} = G^{r}_{\ r} = G^{\theta}_{\theta} = G^{\phi}_{\phi}$. The analytical expressions for the effective energy density and pressure are given by \cite{Sengupta:2023yof}:
\begin{eqnarray}
\rho_e(r) &=& \rho\left(1-\frac{\rho}{\rho_c}\right),\label{rhoeff}\\
p_e(r)&=& p-\rho\left(\frac{2p+\rho}{\rho_c}\right),\label{peff}
\label{edensity}
\end{eqnarray}
where $\rho_c$ represents the critical density in the context of LQC. It is noteworthy that these effective quantities found application within the domain of traversable wormholes in the braneworld scenario, being integrated alongside other parameters \cite{Sengupta:2021wvi}.

Therefore, moving forward, we will particularly focus on the static and spherically symmetric Morris-Thorne wormhole metric as presented by \cite{Morris:1988cz}:
\begin{equation}\label{metric1}
    ds^2=-e^{2\Phi(r)}dt^2+\frac{dr^2}{1-\frac{b(r)}{r}}+r^2d\Omega_2,
\end{equation}
Here, $\Phi(r)$ represents the redshift function, $b(r)$ is the shape function, and $d\Omega_2=d\theta^2+\sin^2\theta d\phi^2$ denotes the spherical line element. Given the metric ansatz of Eq. \eqref{metric1}, modified Einstein's equations take on their simplest form:
\begin{align}
G_{\ t}^{t} = & \frac{b'}{r^{2}}= 8 \pi \rho_e(r),\label{eq:g00}\\
G_{\ r}^{r} = & -\frac{b}{r^{3}}+ 2\frac{\left(r-b\right)\Phi'}{r^{2}}=  8 \pi p_e(r),\label{eq:grr}\\
G_{\theta}^{\theta} = & G_{\phi}^{\phi} = \left(1-\frac{b}{r}\right)\left[\Phi''+(\Phi')^{2}+\frac{\left(b-rb'\right)}{2r(r-b)}\Phi'+\frac{\left(b-rb'\right)}{2r^2(r-b)}+\frac{\Phi'}{r}\right]=  8 \pi  p_{e}(r),\label{eq:gthetatheta}
\end{align}
The quantities $\rho(r)$ and $p(r)=\omega \rho(r)$ which entry in the effective densities and pressures are described by Eq. (\ref{casimirdenspress}), and will be regarded as the sources for the new wormhole solutions investigated here. 

Another feature to be studied in the traversable wormhole solutions that we will obtain later is the conservation equation, which for an isotropic fluid is given by
\begin{equation}\label{TOV}
    -\frac{dp_e}{dr}-\Phi'(\rho_e+p_e)=0,
\end{equation}
in the context of the modified Einstein's equations under inspection.

\subsection{Wormhole solution}

To determine the wormhole solution given the Casimir energy density and isotropic pressure with GUP correction, we first solve for the shape function using the modified Einstein equation for the effective energy density. Then, we use the conservation equation to find the redshift function. The apparent inconsistency with the remaining Einstein equations is overcome since these equations do not provide independent constraints, making them redundant. Additional constraints often emerge from the need to ensure the regularity of the redshift function and analyze junction conditions. However, our use of the conservation equation will reveal that imposing regularity on the redshift function is unnecessary.

\begin{figure}
\centering
   \includegraphics[scale = 0.63]{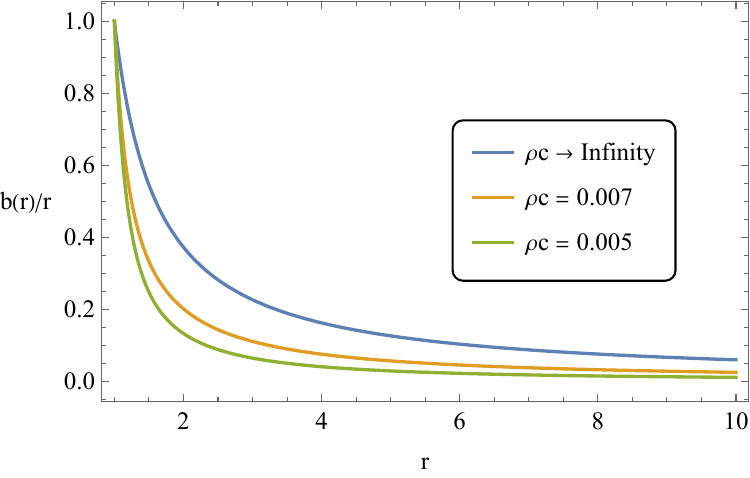}
   \includegraphics[scale = 0.63]{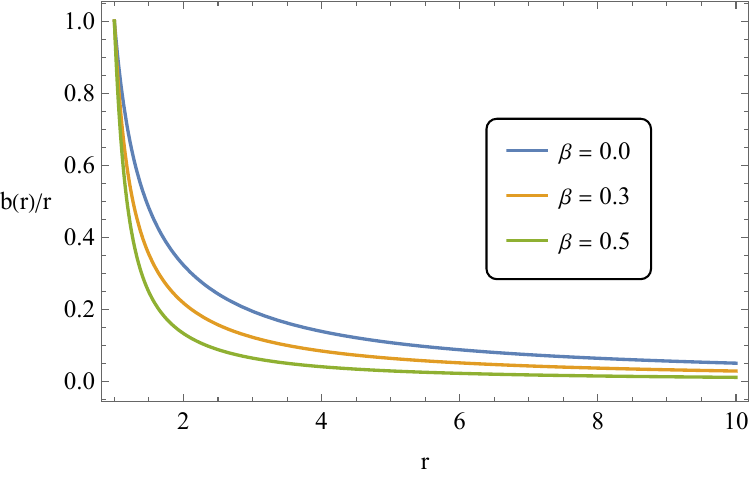}
\caption{Left panel: Plot of $b(r)/r$, for some values of $\rho_c$, considering $\beta=0.5$ and $r_0=1$. Right panel: The same function for some values of $\beta$, considering $\rho_c=0.005$ and $r_0=1$.}
\label{Fig1}
\end{figure}

Thus, from Eqs. (\ref{casimirdenspress}), (\ref{rhoeff}), and (\ref{eq:g00}), we find that the wormhole shape function is given by
\begin{eqnarray}
    b(r)=r_0&-&\frac{8 \pi  k}{ r_0}\left(1+\frac{5\beta }{9 r_0^2}+\frac{ k}{5 \rho_c r_0^4}+\frac{10 k \beta }{21\rho_c r_0^6}+\frac{ 25 k\beta^2 }{81 \rho_c r_0^8}\right)\nonumber\\
    &+&\frac{8 \pi  k}{r}\left(1+\frac{5\beta }{9 r^2}+\frac{ k}{5 \rho_c r^4}+\frac{10 k \beta }{21\rho_c r^6}+\frac{ 25 k\beta^2 }{81 \rho_c r^8}\right).
    \label{b(r)}
\end{eqnarray}
where we have taken into account the boundary condition $b(r_0)=r_0$ to determine the integration constant. According to the prescription found in Refs. \cite{Morris:1988cz, Morris:1988tu}, the shape function determines the geometry of the wormholes and it must adhere to the other following criteria: (i) For radial distances $r > r_0$, the ratio of the shape function to the radial distance must be less than unity, expressed as $b(r)/r < 1$. (ii) The flaring-out condition $b(r) - b'(r) r > 0$ to ensure the throat of the wormhole is the smallest size of the whole structure. (iii) The derivative of the shape function concerning the radial distance at the throat must be less than unity, i.e., $b'(r)_{r=r_0} < 1$. Finally, (iii) implies a minimum size for the throat, which, in turn, minimizes the amount of exotic matter required at the throat to violate the NEC.
In Figure \ref{Fig1}, we can verify that i) is satisfied, as well as ii), since the derivative of the shape function at the throat is always negative. In the left panel, we vary $\rho_c$ and fix $\beta$. In the right panel, $\beta$ is variable and $\rho_c$ fixed.

\begin{figure}
    \centering
    \includegraphics[scale=0.63]{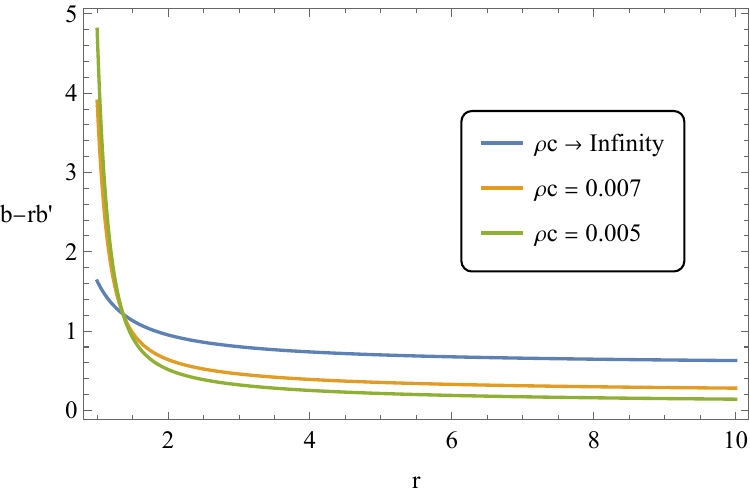}
    \includegraphics[scale=0.63]{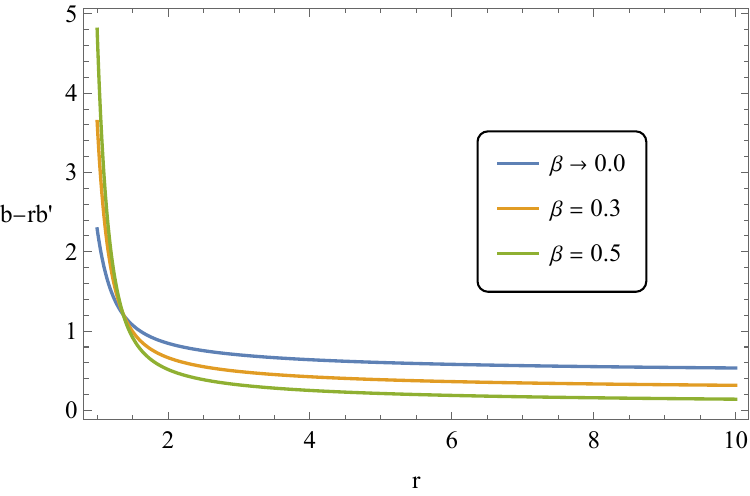}
    \caption{Plot of the flaring-out condition as a function of $r$ where they are positive everywhere to ensure the throat is the narrowest part of the whole structure.}
    \label{fig_flaringout}
\end{figure}

\begin{figure}
    \centering
    \includegraphics[scale = 0.62]{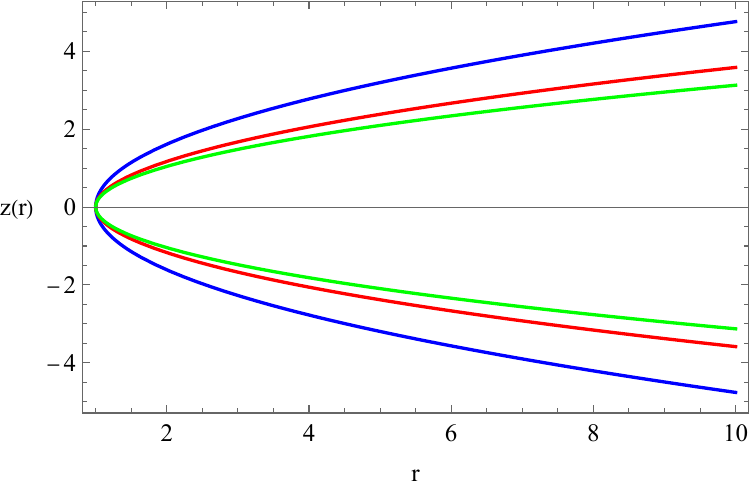}
    \includegraphics[scale = 0.62]{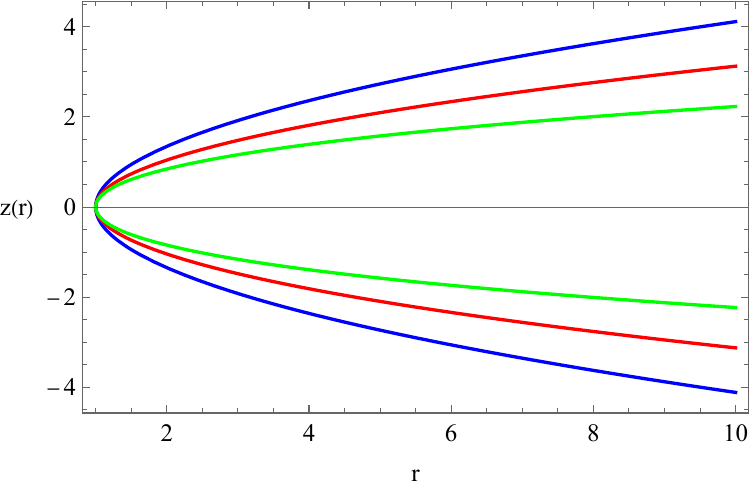}
     \includegraphics[scale = 0.45]{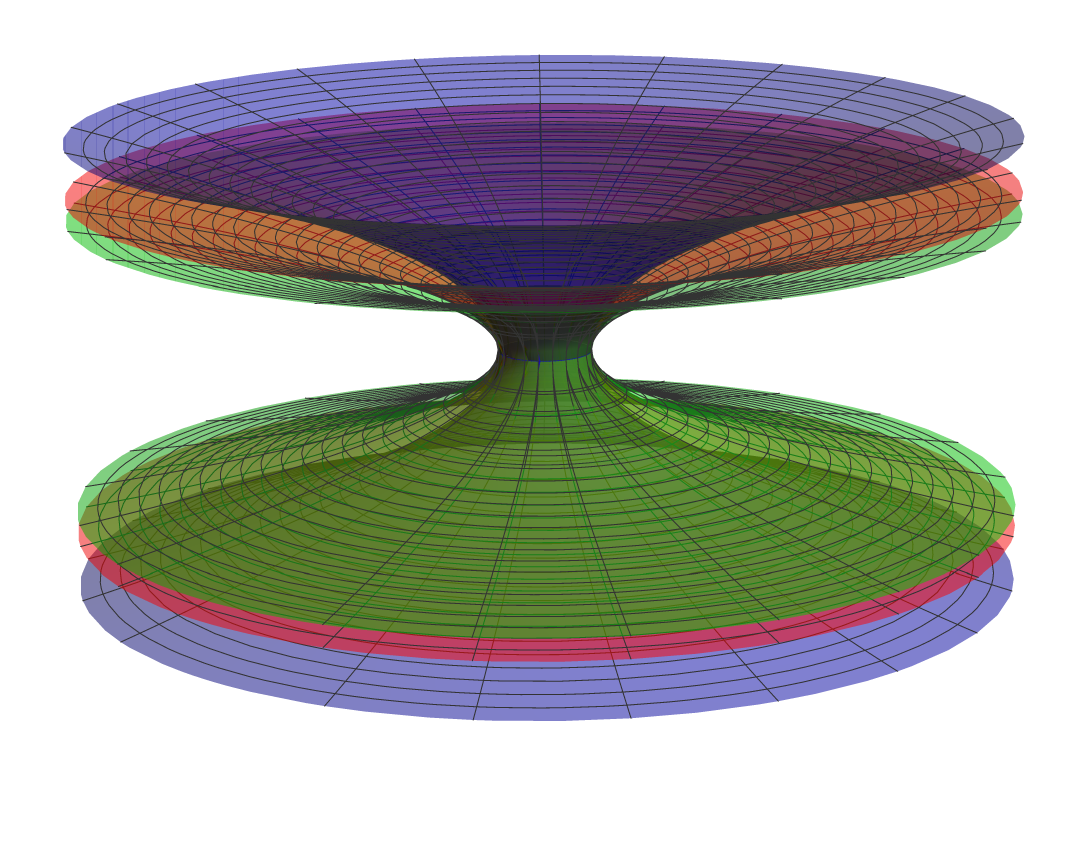}
    \caption{Top, left panel: Plot of $z(r)$, for some values of the critical density: $\rho_c\to \infty$ (i.e., no LQC correction, in blue), $\rho_c=0.007$ (red), and $\rho_c=0.005$ (green), with $\beta=0.3$ and $r_0=1$. Top, right panel: Function $z(r)$ for some values of GUP parameter: $\beta=0.0$ (no GUP correction, in blue), $\beta=0.3$ (red), and $\beta=0.5$ (green), with $\rho_c=0.005$ and $r_0=1$. Bottom: The 3D embedding diagram of some traversable Casimir GUP-corrected wormholes in LQC, with $\rho_c \to \infty$ (purple), $\rho_c = 0.007$ (orange), and $\rho_c = 0.005$ (green), for $\beta=0.3$ and $r_0=1$.}
    \label{profilesz(r)}
\end{figure}
In Figure \ref{profilesz(r)}, we plot some profiles of embedding diagrams for the LQC-Casimir GUP corrected wormholes, which are generated from the mapping of the metric spatial sector in cylindrical coordinates at the equatorial plane, via
\begin{equation}
   dr^2+r^2d\phi^2+dz^2=\frac{dr^2}{1-\frac{b(r)}{r}}+r^2d\phi^2\Rightarrow z(r)=\int_0^{r}\left[\frac{b(u)/u}{1-b(u)/u}\right]^{1/2}du.
\end{equation}
Note that as the value of $\rho_c$ decreases, in the left panel, ($\beta$ increases, in the right panel), indicating a more evident manifestation of quantum gravity, the inclination towards the wormhole throat becomes less pronounced, approaching asymptotic flatness. The 3-D embedding diagram indicates the same behavior when we vary $\rho_c$ and fix $\beta$.

Regarding the redshift function, from Eq. (\ref{TOV}), we find
\begin{equation}
   -g_{tt}(r)= e^{2\Phi(r)}=\left(\frac{r}{r_0}\right)^{\frac{12(1+2\omega)}{1+\omega}}\left(\frac{5\beta+3r_0^2}{5\beta+3r^2}\right)^{\frac{2\omega}{1+\omega}}\left(\frac{10\beta k+6k r_0^2+3\rho_cr_0^6}{10\beta k+6k r^2+3\rho_cr^6}\right)^2,\label{redshifteq}
\end{equation}
The integration constant was chosen to ensure $e^{2\Phi(r_0)}=1$. It's important to note that while the resulting solution remains regular within the domain of $r$, the solution expressed in Eq.(\ref{redshifteq}) lacks asymptotic flatness, although, as we will see, the curvature scalars vanish at infinity. Therefore, we must apply junction conditions such that beyond a certain radius $r=R$, the spacetime transitions to the Schwarzschild vacuum solution. This transition and its implications will be discussed in detail in the next section.

\subsection{Energy conditions}

\begin{figure}
    \centering
    \includegraphics[scale = 0.6]{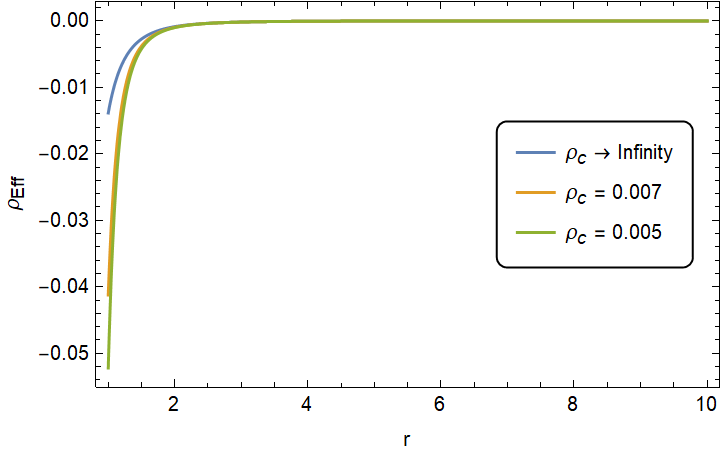}
    \includegraphics[scale = 0.6]{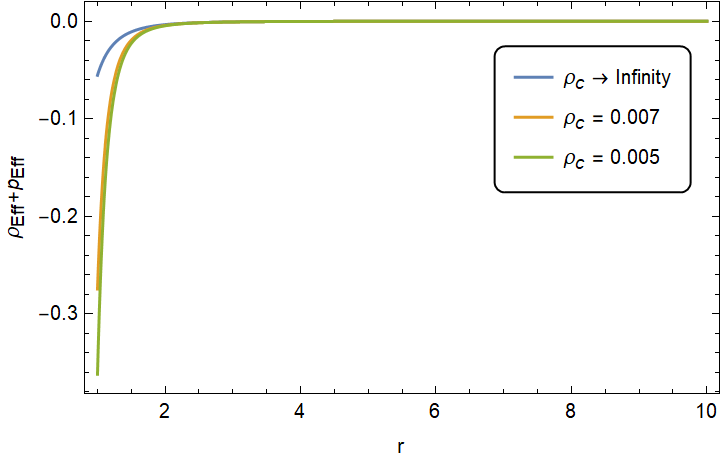}
    \includegraphics[scale = 0.6]{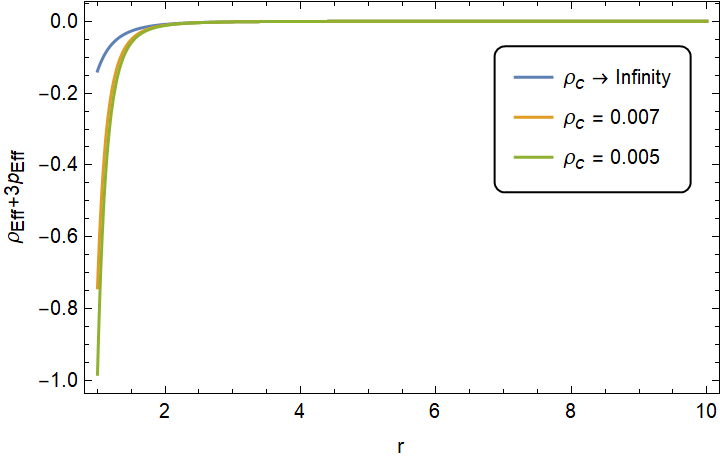}
    \caption{The profiles of the effective energy density $\rho_e$ (top left), $\rho_e + p_e$ (top right), and $\rho_e + p_e$ (bottom) with $\beta = 0.01$ and $\rho_c = 0.005, 0.007, \infty$ presents the violation of the traversable wormhole on the energy conditions i.e. weak, null, and strong energy conditions; however, an increase of $\rho_c$ reduces the amount of the exotic matter and violation area of the energy conditions.}
    \label{fig_en_conds}
\end{figure}

\begin{figure}
    \centering
    \includegraphics[scale = 0.6]{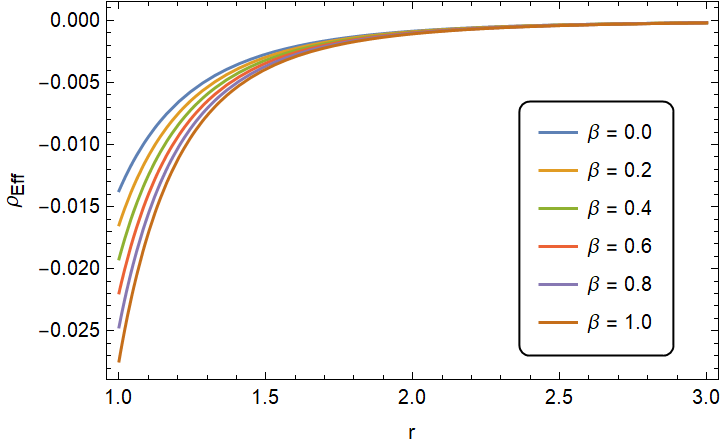}
    \includegraphics[scale = 0.6]{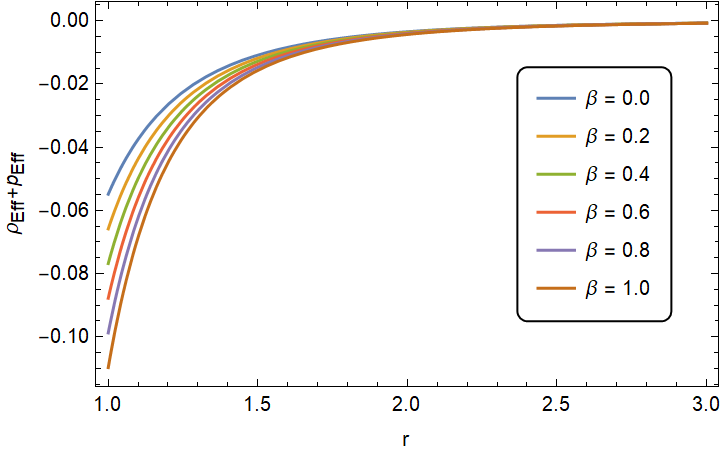}
    \includegraphics[scale = 0.6]{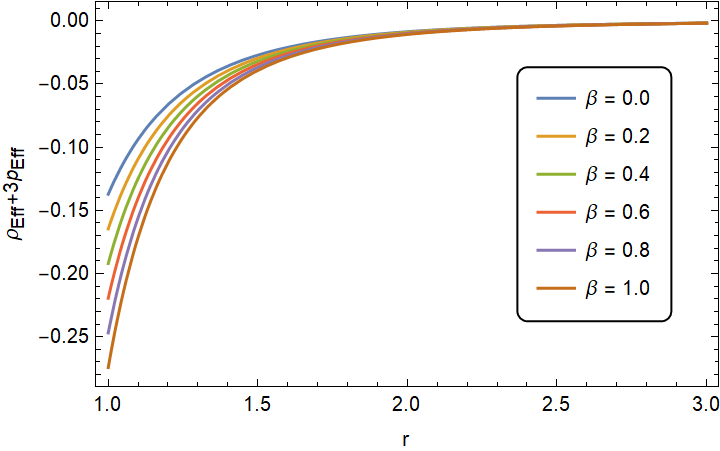}
    \caption{The profiles of the effective energy density $\rho_e$ (top left), $\rho_e + p_e$ (top right), and $\rho_e + p_e$ (bottom) with $\rho_c \rightarrow \infty$ and $\beta \in [0,1]$.}
    \label{fig_en_conds_varyBeta}
\end{figure}

To ensure the existence of solutions within the framework of GR and its modifications, one considers realistic sources for the energy-momentum tensor. We apply the coordinate-independent description of the energy conditions, i.e., null (NEC), weak (WEC), and strong (SEC), on the Casimir traversable wormholes with the GUP correction in the LQC scenario. Thus, we consider the isotropic wormhole and the effective energy-momentum tensor given in Eq.~(\ref{en_mo_tensor}).
\begin{enumerate}
    \item NEC is defined as $T_{\mu \nu} k^{\mu} k^{\nu} > 0$ for all null vectors $k^{\mu}$ which provides the condition 
    \begin{eqnarray}
        \rho_e + p_e \geq 0.
    \end{eqnarray}
    \item WEC is defined as $T_{\mu \nu} t^{\mu} t^{\nu} > 0$ for all timelike vectors $t^{\mu}$ which provides the conditions
    \begin{eqnarray}
        \rho_e &\geq& 0, \\
        \rho_e + p_e &\geq& 0
    \end{eqnarray}
    \item SEC is defined as $\left( T_{\mu \nu} - \frac{1}{2} T g_{\mu \nu} \right) t^{\mu} t^{\nu} \geq 0$ which provides the conditions
    \begin{eqnarray}
        \rho_e + 3 p_e &\geq& 0, \\
        \rho_e + p_e &\geq& 0
    \end{eqnarray}
\end{enumerate}

We can observe a violation of the energy conditions, particularly in the vicinity of the wormhole's throat, consistent with the findings of Li and Zhu \cite{Li:2008sw}. We also calculate the quantities $\rho_e, \rho_e + p_e$ and $\rho_e + 3 p_e$ for investigating the energy conditions' violation of these wormholes. 

In fig.~\ref{fig_en_conds}, we find that the critical energy density $\rho_c$ could reduce the violation level on NEC, WEC and SEC around the throat of the traversable wormholes as shown that $\rho_e, \rho_e + p_e$ and $\rho_e + 3 p_e$ have the less negative at the throat with $\rho_c \rightarrow \infty$.

We consider the effect of GUP by varying \(\beta \in [0,1]\). Higher \(\beta\) values lead to greater violations of the NEC, WEC, and SEC around the wormhole throat. Similarly, decreasing \(\rho_c\) increases the violation of energy conditions near the throat, indicating that stronger quantum gravity effects accentuate these violations.

\subsection{Stability of the solutions}

\begin{figure}[h]
    \centering
    \includegraphics[height=6.3cm, scale = 0.20]{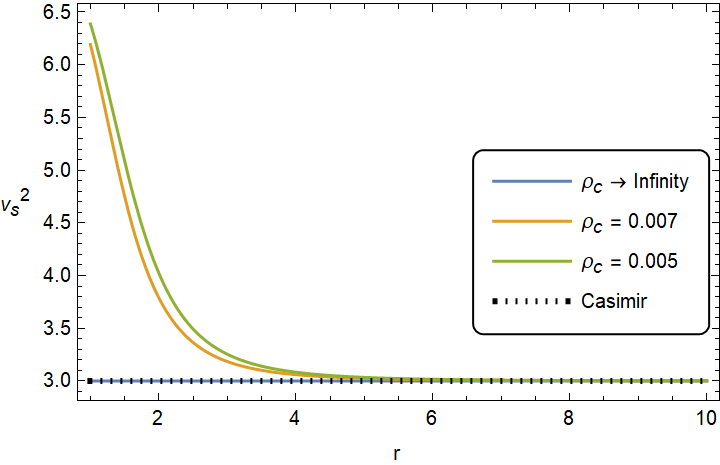}
    \includegraphics[scale = 0.50]{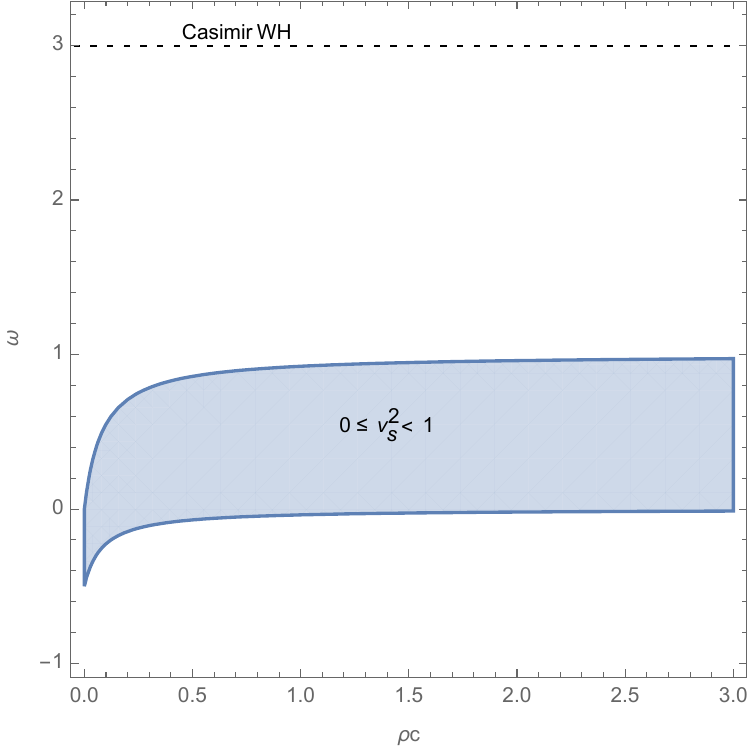}
    \caption{Left panel: Square of the sound speed associated with the source of the Casimir wormholes ($\omega=3$), with the parameter set $\beta = 0.01$, $r_0=1$ and some values of $\rho_c$. Right panel: Space parameter ($\rho_c,\omega$), indicating the region where the wormhole is stable and $v_s<1$, near the throat, for $r_0=1$ and $\beta=0.3$.}
    \label{fig_vsq}
\end{figure}
We will now investigate the stability of the obtained wormhole solutions by analyzing the squared sound velocity of the fluid along the radial direction \cite{Capozziello:2022zoz}. This quantity, denoted as $(v_s)^2$, can be calculated as follows:
\begin{equation}
  \label{vsound}
    (v_s)^2=\frac{d p_e}{d\rho_e}=\frac{d p_e/dr}{d\rho_e/dr}.
\end{equation}
The stability of the wormhole is assured if $(v_s)^2 > 0$, and it must satisfy $(v_s)^2 < 1$ to be physically meaningful. In the left panel of Figure \ref{fig_vsq}, we illustrate this quantity's radial dependence for the strict Casimir wormhole ($\omega=3$), while varying the critical density parameter $\rho_c$ and holding the remaining parameters constant. Despite the wormhole's stability, it's noteworthy that $v_s^2>1$. In the right panel of the same figure, we relax the constraint on pressure to be solely Casimir, thus mapping out the parameter space ($\omega, \rho_c)$ where both stability and physicality conditions are satisfied. It is worth highlighting that for significant quantum corrections stemming from LQC ($\rho_c\to 0$), these conditions are only fulfilled for exotic matter within a range of negative $\omega$.

\subsection{Curvature scalars}

\begin{figure}
    \centering
    \includegraphics[scale = 0.62]{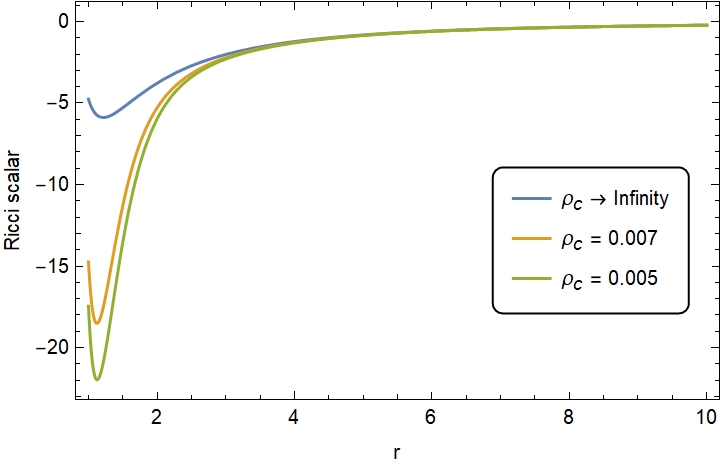}
    \includegraphics[scale = 0.62]{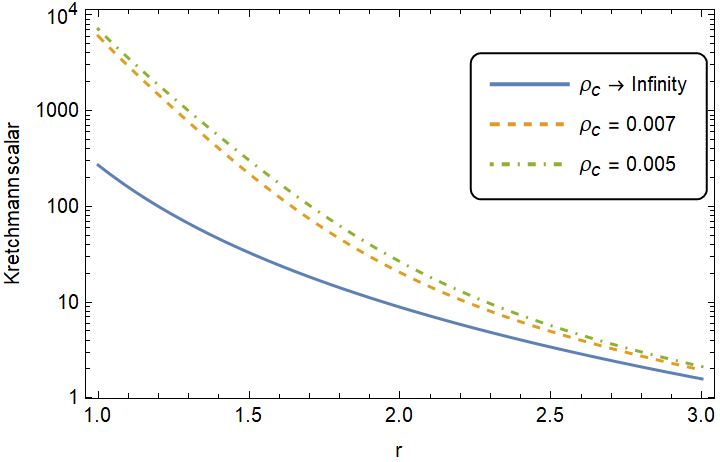}
    \caption{The Ricci and Kretschmann scalars (presented in semi-log scale) of the traversable wormholes with the parameter set $r_0 = 1, \beta = 0.01, \omega = 3$ (Casimir) and $\rho_c = \{0.005, 0.007, \infty\}$.}
    \label{fig_scalar_vary_rho}
\end{figure}
\begin{figure}
    \centering
    \includegraphics[scale = 0.62]{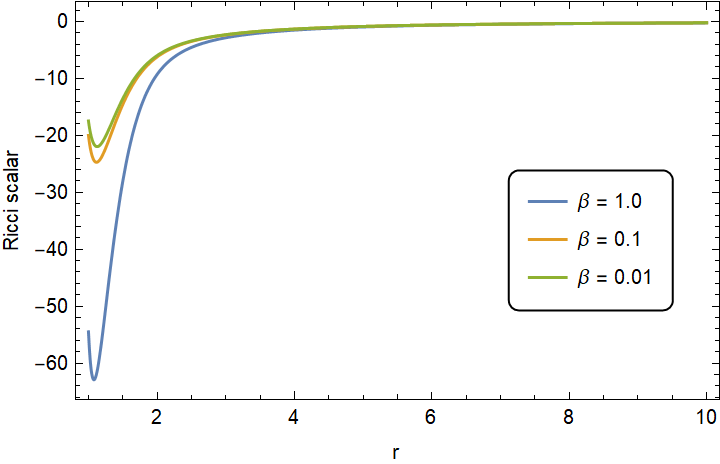}
    \includegraphics[scale = 0.62]{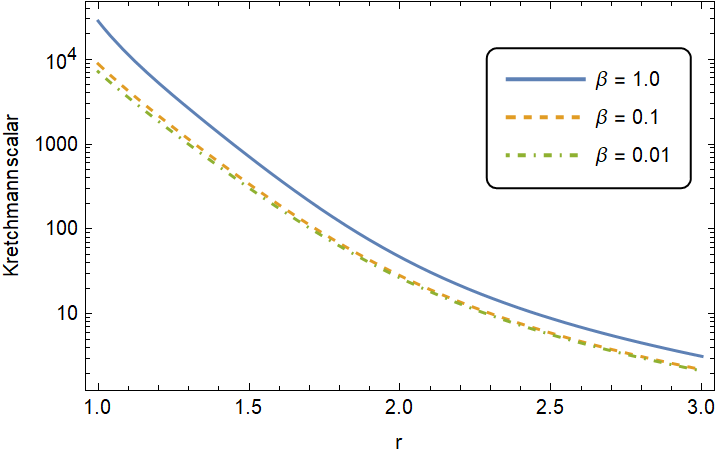}
    \caption{The Ricci and Kretschmann scalars (presented in semi-log scale) of the traversable wormholes with the parameter set $r_0 = 1, \rho_c = 0.005, \omega = 3$ (Casimir) and $\beta = \{ 0.01, 0.1, 1.0\}$.}
    \label{fig_scalar_vary_beta}
\end{figure}

The geometry of the wormhole is one of the major factors contributing to the possibility of traveling. The investigation of the whole structure for the singularities by considering the divergence from Eq.~(\ref{metric1}) is not sufficient due to the coordinate-dependent metric tensor. To find the singularities, we present the coordinate-independent quantities; Ricci and Kretschmann scalars as shown in Fig. \ref{fig_scalar_vary_rho} and \ref{fig_scalar_vary_beta}. 

Considering the effect of the central energy density on the structure, we find that the Ricci scalar is lowest near its wormhole's throat ($r = r_0 = 1$) but never diverges as in the left panel of Fig. \ref{fig_scalar_vary_rho}. The minimum of the Ricci scalar increases as the central energy density increases. Moreover, the negativity effect reduces as the distance increases from the throat. On the right panel of Fig. \ref{fig_scalar_vary_rho}, the value of the Kretschmann scalar drops rapidly and there is no divergence as shown in the semi-log plot. Its maximum is exactly at the throat of the wormhole. The higher central energy density decreases the maximum of the Kretschmann scalar at the throat. As distance from the throat increases, the Kretschmann scalar also decreases.

Next, we consider the effect of the constant of GUP correction $\beta$ on the structure. On the left panel of Fig.~\ref{fig_scalar_vary_beta}, the lowest value of the Ricci scalar is near the throat of the wormhole and it decreases as $\beta$ increases. Like the previous case, the negativity effect fades as the distance from the throat increases. On the right panel of Fig.~\ref{fig_scalar_vary_beta}, an increase in $\beta$ causes an increase of the Kretschmann scalar at the throat. The value rapidly reduces as the distance increases. 

We can conclude that the structure of the traversable wormhole with GUP correction in the context of LQC does not have the singularity which confirms that the travel by this structure is possible. Besides this, the quantum effects due to these corrections increase the absolute value of the curvatures near the wormhole throat.

\section{Junction Conditions}

Given that we must have two spacetimes separated by a surface with radius \(R\), it is essential to establish the junction conditions to ensure the continuity of spacetime and to prevent the occurrence of singularities on this surface. Consequently, we expect the metric to be continuous across the junction surface. To achieve this, we will apply the Israel junction conditions, which involve calculating the surface energy density and surface tension \cite{Israel:1966rt}. These conditions are crucial for maintaining the smoothness and physical plausibility of the wormhole structure.

To calculate the surface energy density and the surface tension, we need to obtain the intrinsic energy-momentum tensor, $S_{ij}$, which generally, is given by the Lanczos equation \cite{Lanczos:1924bgi}:
\begin{equation}
    {S^i}_j=-{\kappa^i}_j+{\delta^i}_j\kappa^k_k,
\end{equation}
where $\kappa_{ij}$ represents the discontinuity in the extrinsic curvature across the surface and is written as
\begin{equation}
    \kappa_{ij}=\kappa_{ij}^+-\kappa_{ij}^-.
\end{equation}
The extrinsic curvature is expressed as
\begin{equation}
    \kappa_{ij}^\pm=- n_\nu^\pm\left.\left[\frac{\partial^2 X_\nu}{\partial\xi^i\partial\xi^j}+\Gamma^\nu_{\alpha\beta}\frac{\partial X^\alpha}{\partial\xi^i}\frac{\partial X^\beta}{\partial\xi^j}\right]\right|_S,
\end{equation}
with $n_\nu^\pm$ being a unit vector with $n^\nu n_\nu=1$ and is written as
\begin{equation}
    n_\nu^\pm=\pm \left|g^{\alpha\beta }\frac{\partial f}{\partial X^\alpha}\frac{\partial f}{\partial X^\beta}\right|^{-1/2}\frac{\partial f}{\partial X^\nu},
\end{equation}
while the intrinsic coordinate at the surface of the wormhole is represented by $\xi^i$. The intrinsic coordinate also satisfies the parametric equation $f\left(x^\alpha(\xi^i)\right)=0$.

\begin{figure}
    \centering
    \includegraphics[scale = 0.62]{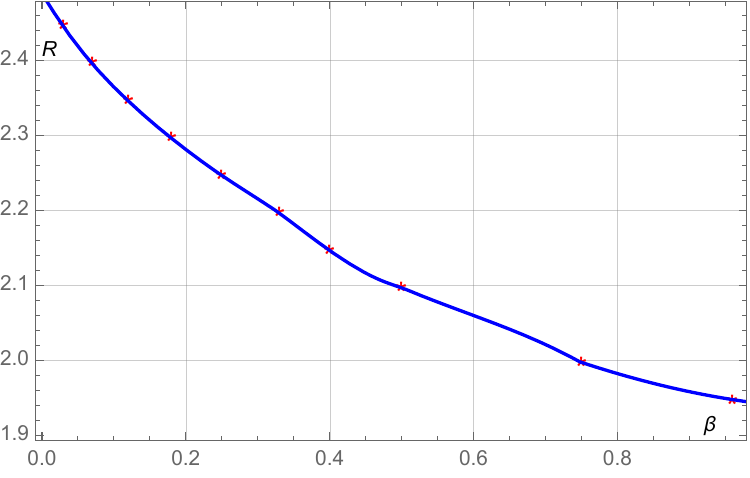}
    \includegraphics[scale = 0.63]{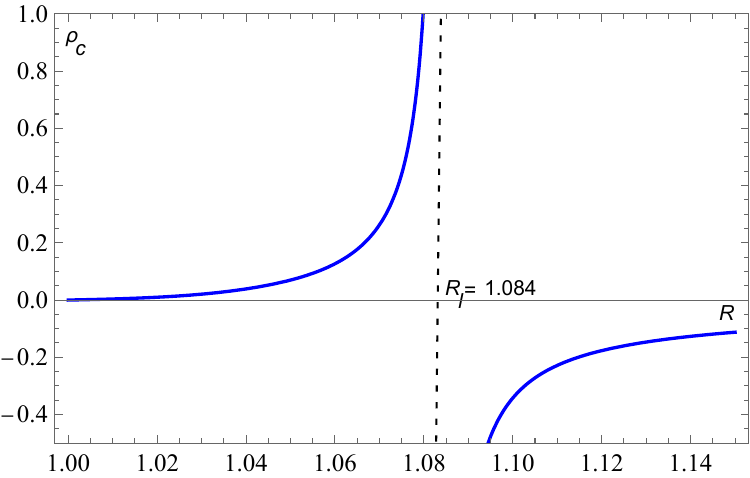}
    \caption{Left panel: Radius of spacetime juntion, $R$, as a function of $\beta$, for $\omega=-1/3$ and $r_0=1$, considering the matching of the three junction conditions. Right panel: Critical density $\rho_c$ as a function of $R$, for $r_0=1$, $\beta=0.5$, and $\omega=3$ (Casimir WH), considering only the first two junction conditions. }
    \label{Fig1junction}
\end{figure}

For a spherically symmetric spacetime, the energy-momentum tensor can be expressed in terms of the surface energy density, $\Sigma$, and the surface tension, $\mathcal{P}$, as \cite{Lobo:2005yv,Lobo:2004id}:
\begin{equation}
    {S^i}_j = \text{diag}\left(-\Sigma,  \mathcal{P},  \mathcal{P}, \mathcal{P} \right),\label{Intrisic_SET}
\end{equation}
where
\begin{eqnarray}
    &&\Sigma = -\frac{2}{R}\left(\sqrt{1-\frac{2M}{R}}-\sqrt{1-\frac{b(R)}{R}} \right),\label{surface_energy}\\
    &&\mathcal{P}= \frac{1}{R}\left[ \frac{1-\frac{M}{R}}{\sqrt{1-\frac{2M}{R}}}-\left(1+R\Phi'(R)\right)\sqrt{1-\frac{b(R)}{R}}\right],\label{surface_tension}
\end{eqnarray}
where $M$ is the mass of the Schwarzschild solution, interpreted here as the proper mass of the wormhole, and $r=R$ is where the junction conditions will be valid. These conditions will combine thus the Casimir wormhole solution with the Schwarzschild one. The derivative of redshift function $\Phi'=d\Phi(r)/dr$ at this radius is
\begin{equation}\label{derivativePhi}
    \Phi^{'}(R)=\frac{6 \left(5 \beta+2 R^2\right) \left[2 k (2 \omega+1) \left(5 \beta+3 R^2\right)+3 \rho_c R^6 \omega\right]}{R (\omega+1) \left(5 \beta+3 R^2\right) \left(10 k \beta+6 k R^2+3\rho_c R^6\right)}.
\end{equation}
At the junction surface, the surface density and the surface tension must be null, in such a way that we obtain \cite{Sengupta:2023ysx}:
\begin{eqnarray}
    && \sqrt{1-\frac{2M}{R}}=\sqrt{1-\frac{b(R)}{R}},\label{cond_1}\\
    && \frac{1-\frac{M}{R}}{\sqrt{1-\frac{2M}{R}}}=\left(1+R\Phi'(R)\right)\sqrt{1-\frac{b(R)}{R}}.\label{cond_2}
\end{eqnarray}
The condition \eqref{cond_1} is equivalent to impose $\left.g_{rr}\right|_{int}=\left.g_{rr}\right|_{ext}$. From it one we have that $b(R)=2M$. Substituting this in (\ref{cond_2}) and using Eqs. (\ref{b(r)}) and (\ref{derivativePhi}), we can find a very involved expression for $\rho_c$ as a function of $r_0$, $\beta$, $\omega$, and $R$. A third junction condition arises from matching the derivatives of \( g_{rr} \) between the wormhole and Schwarzschild spacetimes at \( r = R \). Considering these three conditions, it can be shown that for \( \omega = 3 \), only the first two are satisfied. However, for a range of negative values of \( \omega \), all three conditions are met. In left panel of Figure \ref{Fig1junction}, we fit the curve that best matches the points \((\beta, R)\) obtained from the three junction conditions for \( \omega = -1/3 \) and \( r_0 = 1 \). It is evident that for lower values of \( \beta \), the junction radius \( R \) is larger. For instance, for \( \beta = 0.073 \), \( R = 2.4 \). Hence we calculate \( \rho_c \approx 0.0009 \). Therefore, while the quantum correction due to GUP is small, the correction related to LQC is very high (\( \rho_c \to 0 \)). thus, in this case, the wormhole spacetime extends compared to the Schwarzschild spacetime due to the differing impacts of quantum corrections from GUP and LQC. Large GUP corrections minimally alter the spacetime, leading to slight extensions of the wormhole. In contrast, significant LQC corrections, which introduce a fundamental maximal density, prevent singularities and drastically modify spacetime, resulting in a much more pronounced extension of the wormhole structure.

On the other hand, when we consider solely the conditions involving the tension and pressure on the junction surface, we have a wider variety of scenarios. In Figure \ref{Fig1junction}, right panel, we illustrate the critical density as a function of \(R\) for \(r_0 = 1\) and \(\omega = 3\). Notably, for these strict Casimir wormholes, the possible junction conditions cannot be maintained far from the wormhole throat. In this case, the distance limit is approximately \(R_l \approx 1.084\). For lower values of \(\beta\) (i.e., the lesser the correction due to the Generalized Uncertainty Principle), this distance can be shown to be slightly greater. 

\begin{figure}
    \centering
    \includegraphics[scale = 0.55]{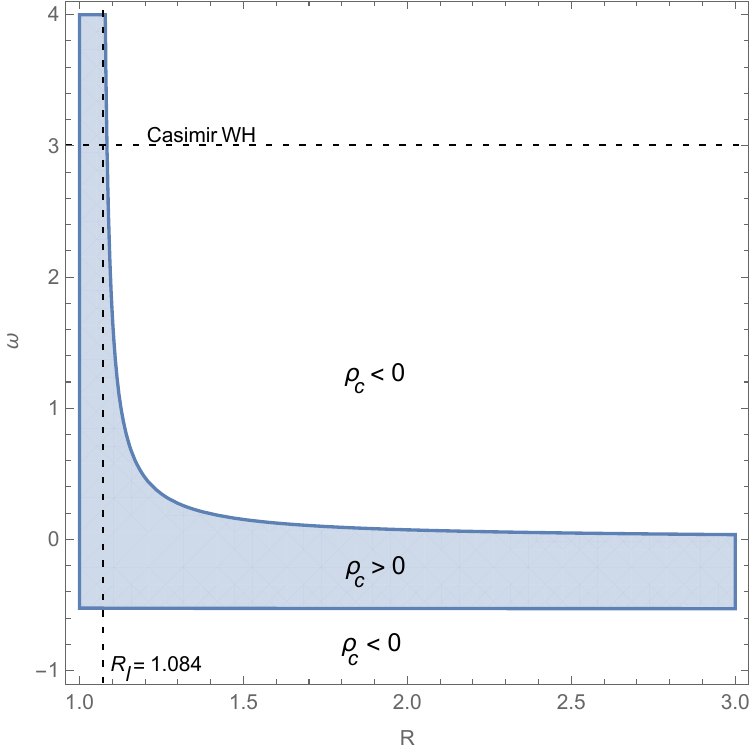}
    \caption{Parameter space ($R,\omega$), highlighting the regions where $\rho_c>0$, considering that only the first two junction conditions are satisfied. Set parameters: $\beta=0.5$, $r_0=1$.}
    \label{ploterreomega}
\end{figure}

In Fig. \ref{ploterreomega}, we depict the parameter space \((\omega, R)\), highlighting the allowed regions where \(\rho_c > 0\). It is noteworthy that for \(-\frac{1}{2} < \omega < 0\), i.e., deformed Casimir wormholes with exotic matter, the junction position can occur at any point, even very far from the throat. 
Thus, in the context of a Casimir wormhole corrected by GUP, with $\omega=3$, a larger parameter $\beta$ corresponds to stronger quantum corrections, which can counteract the gravitational collapse and allow the wormhole solution to persist only for short distances before being overtaken by the classical Schwarzschild solution. However, if we consider negative state parameters (exotic matter), the quantum effects can dominate over the classical gravitational collapse, leading to a larger region where the wormhole solution prevails.

\section{Quantity of exotic matter and wormhole visibility}

\begin{figure}
    \centering
    \includegraphics[scale = 0.6]{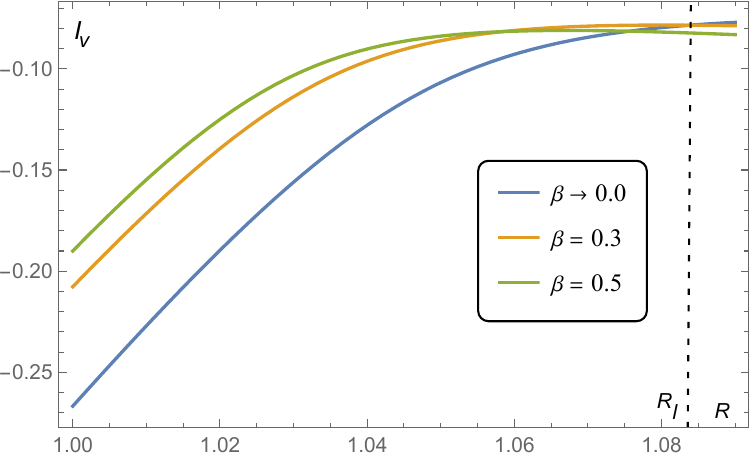}
    \caption{VIQ for some values of $\beta$. Set parameters: $r_0=1$ and $\omega=3$, with $R_l\approx 1.084$.}
    \label{IV}
\end{figure}

We are now going to analyze the Volume Integral Quantifier (VIQ), defined by \cite{Nandi:2004}
\begin{equation}\label{viq}
\mathcal{I}_v=\int_{r_0}^R 4\pi r^2 (\rho_e+p_e)dr,
\end{equation}
where $R$ is the radius of the junction interface. The objective is to obtain the amount of exotic matter necessary to keep its throat open in the scenario under analysis. Thus, we have taken the integral on the effective energy density and pressure, given from LQC combined with the GUP-corrected Casimir quantities. Plugging Eqs. (\ref{casimirdenspress}), (\ref{rhoeff}), and
(\ref{peff}) into (\ref{viq}), we obtain 
\begin{align}
&\mathcal{I}_v=\frac{16 k \pi}{R} - \frac{16 k \pi}{r_0} + \frac{56 k \pi \beta}{9 R^3} - \frac{56 k \pi \beta}{9 r_0^3} + \frac{32 k^2 \pi}{5 R^5 \rho_c} - \frac{32 k^2 \pi}{5 r_0^5 \rho_c} \nonumber\\
&+ \frac{272 k^2 \pi \beta}{21 R^7 \rho_c} - \frac{272 k^2 \pi \beta}{21 r_0^7 \rho_c} + \frac{560 k^2 \pi \beta^2}{81 R^9 \rho_c} - \frac{560 k^2 \pi \beta^2}{81 r_0^9 \rho_c},
\end{align}
for $\omega=3$. The plot in Figure \ref{IV} indicates that, for junction points near the throat of the Casimir wormhole, the amount of exotic matter required is less (more) for higher (lower) quantum corrections due to the GUP, since $\mathcal{I}_v<0$, which is consistent with the violation of the energy conditions. However, for sufficiently distant junction points, approaching the limit where junction conditions are valid in the context of LQC (i.e., where $\rho_c>0$), the amount of exotic matter required is less (more) for lower (higher) values of the GUP parameter. In any case, the further the junction interface is from the throat, the less exotic matter is needed. 
\begin{figure}[h]
    \centering
    \includegraphics[scale = 0.62]{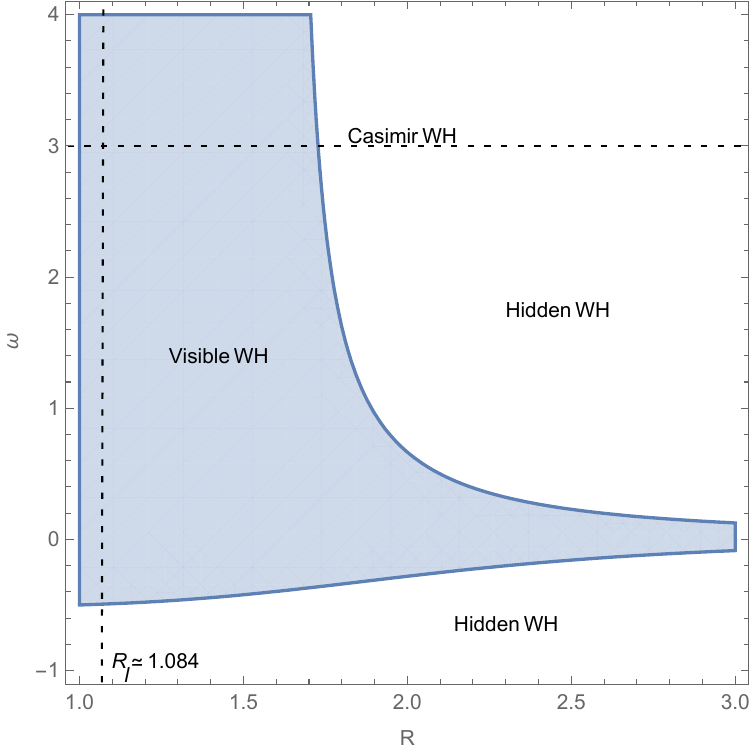}
    \caption{Parameter space ($R,\omega$), highlighting the regions where $2M=b(R)<r_0$, considering that only the first two junction conditions are satisfied. Set parameters: $\beta=0.5$, $r_0=1$.}
    \label{hidvis}
\end{figure}

The presence of a Schwarzschild spacetime beyond the wormhole permits the existence of an event horizon, as we can attribute a mass to the wormhole, \( M = \frac{b(R)}{2} \). Consequently, we can ascertain whether the wormhole is visible or hidden from an external observer, depending on whether \( 2M < r_0 \) or \( 2M > r_0 \), respectively.
Figure \ref{hidvis} illustrates the parameter space $(\omega, R)$, highlighting the regions where the wormhole is either visible or hidden. This analysis assumes that only the first two junction conditions are satisfied. Notably, the strict Casimir wormhole remains consistently visible as long as $R < R_l \approx 1.084$, as previously demonstrated. This consistent visibility within the specified parameter range underscores the distinctiveness of the Casimir wormhole under these conditions.

\section{Conclusions}

In this paper, we have discovered novel static and spherically symmetric traversable isotropic Casimir wormholes with a generic GUP correction within the framework of LQC. We derived the shape function from the modified time component of Einstein's equation and the redshift function from the modified conservation equation, initially considering the linear state equation $p(r)=\omega \rho(r)$
and then specializing to $\omega=3$ for strict Casimir wormholes. Both functions satisfy the criteria for traversability and depend on all involved parameters. Consequently, the violation of energy conditions, embedding diagrams, stability of solutions, and curvature scalars are highly sensitive to the competing quantum gravity effects determined by the GUP and LQC parameters, as demonstrated throughout Section II.

Since the behavior of the redshift function at infinity leads to a solution that is not asymptotically flat, we had to impose junction conditions to integrate the wormhole spacetime with the Schwarzschild one. Our findings revealed that strict Casimir wormholes adhere to only two out of the three junction conditions. The third condition, which allows for a smooth transition from one region to another, is not upheld. In this case, the junction interface is very close to the wormhole throat, according to the right panel of Figure \ref{Fig1junction}. On the other hand, for wormholes sourced by Casimir energy densities and isotropic pressures conforming to equations of state within a specific range of negative $\omega$ values, all junction conditions are met. In other words, this particular exotic matter type enables smooth transitions between the different spacetimes. Consequently, the interface region expands beyond that of the preceding scenario, where the quantum corrections due to GUP are minimal, but the ones due to LQC are maximal, as shown in the left panel of the same figure.

We also have shown that the position of the junction interface is crucial in determining the quantity of exotic matter required to maintain Casimir wormholes with GUP corrections within the context of LQC. If the junction interface is close to the throat, a larger amount of exotic matter is required, consistent with the more pronounced violation of energy conditions at that location. This effect is especially significant when quantum corrections due to GUP and LQC are substantial, as shown in Figure \ref{IV}.

Ultimately, we examined the position of the wormhole solution concerning the event horizon of the Schwarzschild spacetime, as determined by the junction conditions, and evaluated its visibility and accessibility for external observers. Our findings, illustrated in Figure \ref{hidvis}, suggest that the strict Casimir wormhole is fully accessible from the exterior, implying that observing such an object could theoretically be feasible.

\section*{Acknowledgments}
\hspace{0.5cm} CRM thanks the Conselho Nacional de Desenvolvimento Cient\'{i}fico e Tecnol\'{o}gico (CNPq), Grants no. 308268/2021-6. TT is supported by School of Science, Walailak University, Thailand.



\end{document}